# Feature Selection and Intrusion Detection in Cloud Environment based on Machine Learning Algorithms

Amir Javadpour
School of Computer Science and Educational Software, Guangzhou University, Guangzhou, China, 510006
a.javadpour@ut.ac.ir

Sanaz Kazemi Abharian
Department of Computer Engineering, Sharif University of Technology, Tehran, Iran
skazemi@ce.sharif.edu

Guojun Wang
School of Computer Science and Educational Software, Guangzhou University, Guangzhou, China, 510006
csgjwang@gzhu.edu.cn

*Abstract*— Characteristics and way of behavior of attacks and infiltrators on computer networks are usually very difficult and need an expert In addition; the advancement of computer networks, the number of attacks and infiltrations is also increasing. In fact, the knowledge coming from expert will lose its value over time and must be updated and made available to the system and this makes the need for expert person always felt. In machine learning techniques, knowledge is extracted from the data itself which has diminished the role of the expert. Various methods used to detect intrusions, such as statistical models, safe system approach, neural networks, etc., all weaken the fact that it uses all the features of an information packet rotating in the network for intrusion detection. Also, the huge volume of information and the unthinkable state space is also an important issue in the detection of intrusion. Therefore, the need for automatic identification of new and suspicious patterns in attempt for intrusion with the use of more efficient methods (Lower cost and higher performance) is needed more than before. The purpose of this study is to provide a new method based on intrusion detection systems and its various architectures aimed at increasing the accuracy of intrusion detection in cloud computing.

I. INTRODUCTION

Intrusion is a collection of illegal actions that take the system out of safety. A system is secure when it provides confidentiality, integrity and availability for its users. As network-based computer systems play a vital role in today's societies, They are targeted at more attacks and intrusions. In addition to methods of intrusion prevention such as authentication of the user, like the use of encrypted words, the use of firewalls and the protection of information such as encryption, intrusion detection is also used as another wall to protect computer systems[1], [2]. The purpose of the intrusion detection is to detect unauthorized use, Abuse and damage to computer systems and networks by both internal users and external attackers[3].

The unique features of computer networks, and in particular the Internet[4], such as openness[2], comparability and accessibility[2] every time-every Where make it an ideal platform for providing a new generation of online services, such as e-commerce[5], entertainment and Social networks[6]. Business organizations use this platform to sell their products and services, government agencies provide their users with the services they need, and on the other hand users offer their financial and personal information to various sites for the purchase of products and services, presence on the social networks and the use of entertainment sites or the use of government services[7], [8]. The importance of this kind of information has greatly increased the number of attacks and their complexity in cyber space significantly. This has raised security concerns for governments and business organizations. Therefore, researchers have presented variety of solutions to improve the effectiveness of security solutions to detect and prevent attacks. Intrusion detection systems, one of the reactive tools, are widely used as a complementary layer of defense along with these methods. An intrusion detection system is a hardware or software system that examines the input traffic to the network or host in order to detect unconventional attacks and behaviors[2], [9].

The author in [9] proposed a three-layer neural network to detect abuses in the network. The vector attribute used in this scheme includes nine network attributes. The degree of intrusion success with MLP and self-organization map is compared in [10]. The results showed that SOM has higher identification accuracy than NN.

Fuzzy logic is used to deal with the typical description of intrusions[11]. Evolving Fuzzy Neural Network (EFuNN) was developed by the Chavan et al in 2004 to reduce the training time. In this scheme, a combination of controlled and uncontrolled training is used. The experimental results indicate that the use of reduced EFuNN inputs compared to NNs has a more accurate classification for IDS[2], [12]. Method [12] is not applicable for real-time use in detecting network intrusion because the training time is very high. In order to reduce the training time, NN fuzzy logic with NN is used to quickly detect unknown attacks in Cloud [13]. Svm was used to identify intrusions based on limited sample data in [14] so that the data dimensions do not affect the accuracy of the results. Study [15] shows that svm is better than NN in terms of the positive false production rate because NN requires a large amount of tested samples for effective classification while svm requires few parameters while SVM is only used for binary data. Regardless of this, the accuracy of recognition through the combination of SVM improves with other techniques. A Table 1 show summarizes the techniques used in the intrusion domain.

TABLE I. SUMMARY OF TECHNIQUES USED IN IDS.

| Techniques | Features/Benefits | Limits/challenges |
|---|---|---|
| Neural Network | The proper classification of the unstructured network Increasing accuracy with increasing the multiple secret layers | Need more time and samples Low flexibility |
| Fuzzy Logic | Suitable for qualitative features Better flexibility in uncertain problems | Accuracy less than NN |
| SVM | Good accuracy with limited number of samples | Need to preprocessing features |

## II. MODELING AND PROBLEM SOLVING

In this paper, we applied two techniques: Pearson Linear Correlation and mutual information, plugin and irrelevant features are set aside. First the features are reduced using the linear correlation algorithm then for features not selected in the previous section the feature selection algorithm is used using the mutual information criterion. In this method, first the mutual information between each attribute and the class label is computed in the dataset. In fact we rank the features, after sorting features by rank, we consider the feature that has the highest level of mutual information and the mutual information is computed between this feature and other features and the feature which has the least mutual information with the above feature is selected. Finally, the features obtained from both parts are considered as a selectable feature set. In order to evaluate this set of obtained characteristics, the combination of classification algorithms is used. The process of doing the job is in accordance with the flowchart shown in Figure 1.

### A. Feature Selection

Regarding the effect of feature selection on achieving higher accuracy of classifier algorithms, the feature selection in this research is done by using two methods of selecting; method of choosing the feature of linear correlation and the method of choosing feature of the mutual information in the data relating to the prediction of attacks and their type in the data. Correlation-based criteria measure the predictability of a variable's value by another variable. The coefficient of one of the criteria for dependence is classical and we can use it to find a correlation between a feature and a class. If the correlation of the X-feature with the class C is greater than the correlation of the Y-feature with the class C, then the X-feature is superior to the Y-feature. Correlation-based methods are general in the sense that we can use the subset for different classifiers. Also these methods have a low time complexity.

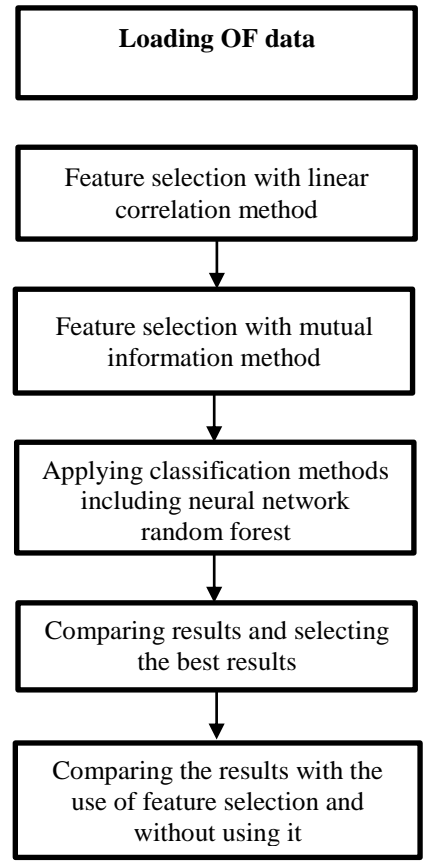

Fig. 1. Flowchart of Proposed Method.

In this paper, Pearson Linear Correlation method was first applied to data[16] . In order to do this, the correlation between the two characteristics was evaluated. The symmetric correlation matrix is obtained from the implementation of this feature selection method.

In the following, the feature selection methods used in this paper are based on information measures. These measures consider the amount of information that is obtained by a feature. For example, the X feature in these methods has priority over the Y feature, if the information obtained from the X feature is greater than the information obtained from the Y feature. The method used in this paper is to use the feature selection method of mutual information according to Study [9].

$$I(X;Y) = H(X) - H(X|Y) = H(Y) - H(Y|X) = -\sum_{y \in Y}\sum_{x \in X} p(x.y) \log \frac{p(x.y)}{p(x)p(y)} \quad (1)$$

The process of work is that by using the joint entropy formula (1), we first compute the mutual information of each feature with the target feature. In order to obtain joint entropy, we need to calculate the probability of the occurrence of different states of each of the features with the objective feature. The concept of shared information is represented by

I(X; Y). If the two features X and Y are closely related, the value of I (X; Y) will be very high. By running this algorithm, some features may be irrelevant and some contain duplicate information. This means that the information obtained by this feature is available in other features. These additional features have led to an increase in the calculation time in the creation of the model. In addition, they can be effective in accuracy and validity of the made classifier.

*B. Classification of attacks*

In order to use different methods of classification and use of the benefits of each, several classification methods were applied to pre-processed data and feature selection. Comparison of different classification methods with a brief description of each is given in Table 2.

TABLE II. COMPARISON OF APPLIED CLASSIFICATION METHODS.

| Method | Description |
|---|---|
| **Neural Network**[8]. | Time to build a network for training data is time consuming. So it is appropriate in cases it can be paid for the cost of this time. |
| **CART[1] Algorithm**[17]. | It is used for discrete variables. A decision tree with two decisions. The Gini index is a suitable selection criterion. |
| **ID3Decision tree algorithms**[18]. | It is very suitable for building a tree with multiple divisions in each node. This algorithm was designed for qualitative variables. The criterion for decision making is the entropy index |
| **Random forest algorithm**[19]. | A tree random forest produces many decisions. Extremely large data can be implemented. It gives estimate of the most important variables in the classification |

*C. Comparison criteria*

In this paper, we will use several evaluation methods and criteria to evaluate the proposed method. The most common method of drawing the matrix is interference (Table 3). The evaluation criteria include the criteria of accuracy, recall and result precision.

TABLE III. VIEW OF THE MATRIX INTERFERENCE.

| Actual/ Predict | Positive | Negative |
|---|---|---|
| Positive | True Positive(TP) | False Negative(FN) |
| Negative | False Positive(FP) | True Negative(TN) |

Each of the matrix elements is as follows:

- TN: Indicates the number of records whose actual category is negative and the proposed system could also correctly identify this category.
- TP: Represents the number of records whose real category is positive and the proposed algorithm also correctly recognizes this category.

[1] Classification and Regression Tree

- FP: Represents the number of records that are negative in their actual data but the proposed algorithm has mistakenly identified it as positive.
- FN: Indicates the number of records whose actual data is positive but the proposed algorithm has put this record in a negative category.

$$Percision = \frac{TP}{TP + FP} \quad (2)$$

$$Recall = \frac{TP}{TP + FN} \quad (3)$$

$$Accuracy = \frac{TP}{TP + FP + TN + FN} \quad (4)$$

III. EVALUATION OF THE RESULTS

The database used in this article is called KDD99 and is available in the UCI Machine Learning Dataset Collection which contains 42 features that represent the last attribute of the data class that is the normalization of a record or attack. The attacks included in the database will include dos and u2r, prop, r21 and their details are discussed in the third chapter. The total number of records is 10000. This database does not contain missing values. Weka software is used to implement these methods. In this section, preprocessing has been applied to the data and feature selection method has been implemented. In this step, the classification algorithms are applied to all the set of features and the set of selected features by using the algorithms described in the previous section and the results are compared with each other.

*A. Feature Selection Results*

In order to use the Pearson correlation property selection method, this method is required as a feature selection method to find the criteria for removing plugin features. The threshold value defined in the initialization of this paper is 0.5. It means if the absolute value of the correlation of the two features is greater than 0.5, one of these features is added and does not need to achieve the goals of this paper. After doing this, the feature selection with the threshold of 0.5, 14 features from the dataset is deleted and the rest of the 25 features are used in the next step.

In the next step, by applying the mutual information feature selection algorithm, we will find additional features that their presence can also have a negative effect on the accuracy of the classifier which leads to an increase in computational time in creating the classification model. The results show that some of these features have almost no shared information with the target class (Figure 2). Therefore, the data class used in the purpose of detecting attacks is irrelevant. In order to achieve the final result of this algorithm, it is necessary to define a threshold based on it, features that do not have the same information as the target feature (data class) are eliminated from the features. The selected initial threshold is 0/001. This means if the shared information value of a feature with the data class is less than value of 0.001, this feature is irrelevant and deleted from the data set. By applying this threshold, 4 other features are also reduced from the set of examined features. 21 final selective features along with the data class will be entered into the next step when the classifier algorithms are implemented.

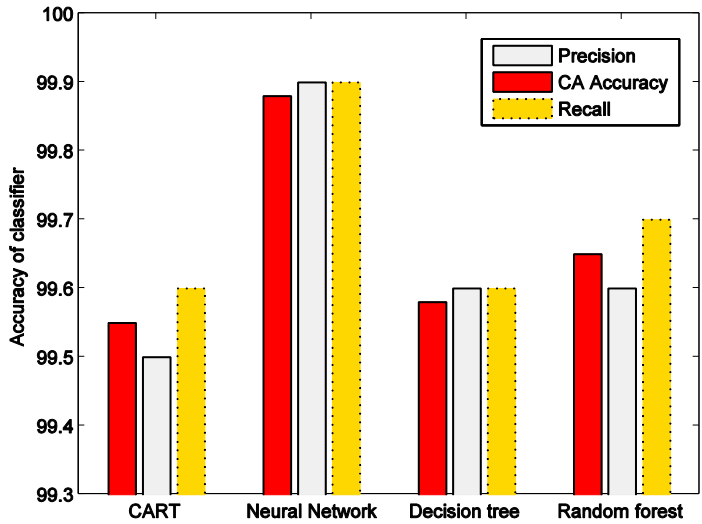

Fig. 2. The accuracy of classification algorithms on the data, along with the proposed feature selection method.

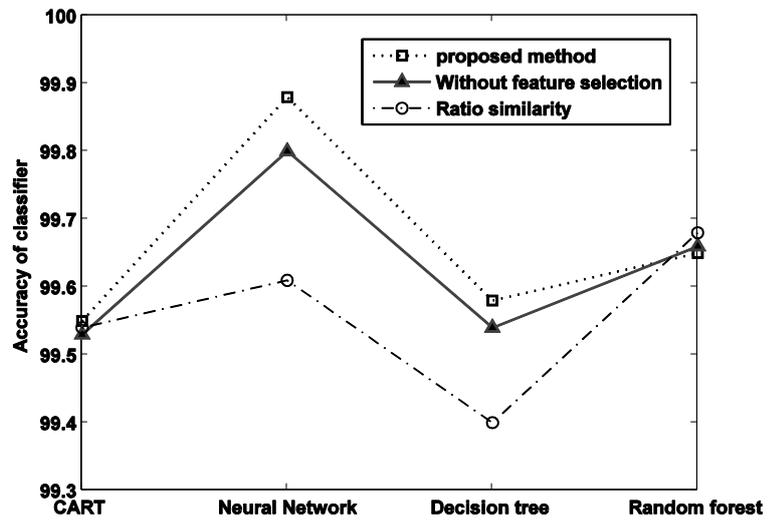

Fig. 3. Accuracy obtained in classification methods by selecting the attribute and without it.

TABLE IV. COMPARISON OF NUMERICAL ACCURACY OF OBTAINED ACCURACY IN CLASSIFICATION METHODS

| Method | Without feature selection | proposed method | Ratio Similarity |
|---|---|---|---|
| CART Algorithm[17]. | 99.53 | 99.55 | 99.54 |
| Neural Network[8]. | 99.80 | 99.88 | 99.61 |
| ID3Decision tree algorithms[18]. | 99.54 | 99.58 | 99.40 |
| Random forest algorithm[19]. | 99.66 | 99.65 | 99.68 |

TABLE V. EVALUATION CRITERIA DERIVED FROM METHODS APPLIED BY FEATURE SELECTION.

| method | Random forest | Decision tree | Neural network | CART |
|---|---|---|---|---|
| CA Accuracy | 99.65 | 99.58 | 99.88 | 99.55 |
| Precision | 99.6 | 99.6 | 99.9 | 99.5 |
| Recall | 99.7 | 99.6 | 99.9 | 99.6 |
| EA error classifier | 0.35 | 0.42 | 0.12 | 0.45 |
| DR Accuracy of detection | 96.93 | 96.09 | 97.91 | 96.4 |
| FAR, False alert rate | 0.0042 | 0.0055 | 0.0031 | 0.0047 |
| AUC | 99.8 | 99.8 | 99.99 | 99.8 |
| Time of building model | 0.78 seconds | 1.26 seconds | 0.74 seconds | 1.16 seconds |

*B. Results of classifier algorithms*

In order to evaluate the results obtained from the implementation of classifier algorithms, we used a 10 fold cross method. Then the results of applying these classification methods are shown in Figure 3. It is observed that the accuracy of the applied methods on data is improved in the case when the feature selection method is used. It is observed that the method of classification of the neural network using the proposed method introduced in this paper has a higher accuracy than other methods. Table 4 and Table 5 show the numerical value of the accuracy obtained in methods with feature selection and without feature selection and Evaluation criteria derived from methods applied by feature selection. In order to better evaluate the proposed method, the accuracy of the classifier algorithms was studied by the method of feature selection of likelihood ratio.

IV. CONCLUSION

In order to provide complete security in Intrusion Detection System, in addition to firewalls and other intrusion prevention devices, other systems called intrusion detection systems are required. In this paper we combined the methods of feature selection of the linear correlation and mutual information. The database is used in this paper is KDD99. Of the 41 features in this database, 21 features are ultimately retained and used in classifier algorithms. Different classification algorithms including decision tree, random forest, CART algorithm and neural network were applied to the data. The results of these methods have been recorded in the previous chapter. The best accuracy obtained in these methods is 99.98% the neural network method. Since the number of records associated with each type of attack is different, one of our suggestions for continuing work is the use of data balancing techniques. Thus, the number of records associated with each type of network status approaches each other. In addition, the method of feature selection can be used in ways such as the INTERACT method in which the relationship between features also influences determining the importance of a feature[20]. Combination information methods such as majority voting can also be used to improve the results achieved in classifications.

V. ACKNOWLEDGEMENT

This work is supported in part by the National Natural Science Foundation of China under Grants 61632009 & 61472451, in part by the Guangdong Provincial Natural Science Foundation under Grant 2017A030308006 and High-Level Talents Program of Higher Education in Guangdong Province under Grant 2016ZJ01.